# Facet-Driven Electronic Modulation in Self-Aligned Phosphorus Nanoarrays

M. Bassotti,[a,*] M. Tallarida,[b] J. Dai,[b] S. Salaverria,[c] P. Angulo-Portugal,[a] L. Fernandez,[a] A. El-Sayed,[a,d] J.E. Ortega,[a,e] A.A. Makarova,[f] Dimas G. de Oteyza,[c] D.Yu. Usachov,[g,h] A. Verdini,[i] and F. Schiller [a,*].

*a) Centro de Física de Materiales CSIC/UPV-EHU – Materials Physics Center, 20018 San Sebastián, Spain*

*b) ALBA Synchrotron Light Source, Cerdanyola del Vallès, 08290 Barcelona, Spain*

*c) Nanomaterials and Nanotechnology Research Center (CINN), CSIC-UNIOVI-PA, 33940 Oviedo, Spain*

*d) Department of Electrical and Computer Engineering, University of California Los Angeles, Los Angeles, CA, 90095 USA*

*e) Departamento de Física Aplicada, Universidad del País Vasco UPV-EHU, San Sebastián, Spain*

*f) Helmholtz-Zentrum Berlin für Materialien und Energie, BESSY II, Albert-Einstein-Str 15, 12489 Berlin, Germany*

*g) St. Petersburg State University, 7/9 Universitetskaya nab., 199034 St. Petersburg, Russia*

*h) Moscow Institute of Physics and Technology, Institute Lane 9, 141701 Dolgoprudny, Russia*

*i) Consiglio Nazionale delle Ricerche – Istituto Officina dei Materiali (IOM), 06123 Perugia, Italy*

** mattia.bassotti@ehu.eus, frederikmichael.schiller@ehu.eus*

## Abstract:

Two-dimensional (2D) materials often require specific substrate terminations for epitaxial stabilization, yet the search for suitable templates has largely focused on low-index metal surfaces, which may not provide the optimal conditions for the growth of new phases. Here, we show that crystal-facet engineering on curved Cu surfaces enables the stabilization, within a single preparation step, of two distinct phosphorus surface phases with different electronic properties. Hexagonal phosphorus adlayer forms on Cu(111) terraces, whereas a previously unreported skewed-square P overlayer is stabilized on Cu(513) facets. By combining complementary microscopy and spectroscopy techniques with theoretical calculations, we show the quasi-semiconducting character of the P-derived electronic states of the skewed-P overlayer, in contrast to the case of hexagonal phosphorene. The coexistence of these two competing structures gives rise to electronic P band modulation overlayer across the substrate while, locally, self-aligned nanoarrays of alternating hexagonal and skewed-square phosphorus terraces form. These results establish crystal-facet engineering as a practical route for discovering and stabilizing emergent 2D phases on high-index substrates.

## Introduction:

The discovery of new two-dimensional (2D) materials and phases increasingly relies on epitaxial growth on crystalline substrates [1,2]. Surface science studies, however, have traditionally focused on low-index metal surfaces [2–4], which provide well-defined and reproducible platforms but may restrict the range of structures that can be stabilized [5]. High-index and vicinal surfaces offer a broader variety of atomic coordinations and geometries, providing alternative stabilization pathways for 2D materials [5]. In this context, crystal-facet engineering [6,7], particularly through curved crystals exposing a continuous distribution of surface orientations [8–11], provides an effective route to explore substrate terminations beyond conventional low-index surfaces.

Growth on vicinal surfaces has already been shown to induce substrate faceting in systems such as graphene and h-BN on (111)-oriented metals [12–14]. In these cases, however, the overlayer largely retains its intrinsic structure while the substrate reorganizes to accommodate it [12–14], limiting the emergence of structurally distinct 2D phases. Phosphorus offers a particularly promising alternative because of its rich polymorphism and the existence of several stable 2D allotropes [15–17]. Among them, phosphorene is an important 2D semiconductor combining a thickness-dependent band gap, high carrier mobility, mechanical flexibility, and large on/off ratios [18–25].

Two-dimensional phosphorus structures have been reported on several substrates, including Pt [26–29], Cu [30,31], Ag [32–34], and Au [35–38] surfaces. A notable example is the phase on Cu(111) previously assigned to blue phosphorene (Blue-P) [39–42]. This phase, however, exhibits metallic character [39], which could arise from strong hybridization with the substrate possibly leading to a $Cu_2P$ surface alloy [43]. Indeed, a $Cu_3P_2$ surface alloy has also been proposed as a precursor to the putative blue-P phase [44]. Thereby, semiconducting phosphorus phases remain comparatively rare, having been obtained, for example, by Silicon intercalation beneath P/Au(111) [45] or in phosphorus chains on Ag(111) [33], while the bottom-up epitaxial growth of black phosphorene remains challenging [46]. These results suggest that exploring non-conventional substrate orientations may provide access to phosphorus phases unavailable on standard low-index surfaces.

Here, we use a curved Cu single crystal spanning the Cu(111)–Cu(513) vicinal range (vicinal angles from 0 to 28.5°) to demonstrate the simultaneous stabilization of two distinct phosphorus phases under identical growth conditions. Besides the previously mentioned hexagonal P structure on Cu(111) terraces [28], we identify a new unreported phase, denoted here skewed-P, associated with a phosphorus-induced Cu(513) facet. Low-Energy Electron Diffraction (LEED), Scanning Tunneling Microscopy (STM), and Photoelectron Diffraction (PED), supported by theoretical simulations, reveal a skewed-square unit cell forming a $1\times1$ structure on the Cu(513) surface. X-Ray Photoemission Spectroscopy (XPS) further identifies a distinct P chemical environment, while Angle-Resolved Photoemission Spectroscopy (ARPES) shows that P-originated bands do not cross the Fermi level, in clear contrast to the case of the metallic P/Cu(111) system. Competition between the two phases across the vicinal region produces a nanometer-scale periodic alternation of phosphorus structures with different electronic properties. Thus, a single phosphorus deposition step combined with substrate faceting provides a simple route toward self-organized 2D nanoarchitectures with spatially modulated electronic properties.

## Results and Discussion:

1. Characterization of the P/Cu(513) system.

Phosphorous deposition on a curved Cu crystal that exposes (111) vicinal surfaces, referred here to as *c*-Cu(645), allowed us to reveal the coexistence of two stable and competing P phases: the debated blue-P growing on the Cu(111) surface [39–44] and a skewed-square P structure growing on Cu(513) terraces. For the sake of simplicity, we focus first on the structural and electronic properties of this new skewed-P superstructure synthesized as a pure phase on a flat Cu(513) crystal, while the more complex double-phase P/*c*-Cu(645) experiment will be discussed at the end.

Figure 1(a) shows the LEED patterns of the pristine and P-covered Cu(513) surfaces. For the pristine Cu(513), the diffraction pattern reveals a skewed-square unit cell, consistent with the mesh formed by the top-layer Cu atoms of the (513) surface [47], depicted as grey spheres in the structural model of Figure 1(b). The bulk Cu atoms (second layer and below) are shown in orange. The LEED pattern of the P-covered surface in Figure 1(a) exhibits qualitatively the same diffraction symmetry, again displaying the skewed-square unit cell, which indicates that P forms a $1\times1$ superstructure with respect to the substrate. However, due to the different scattering response of the P atoms relative to the substrate, the diffraction spots show different relative intensities. The unit cells of both the substrate and the P overlayer are also outlined in Figure 1(b); they share a side length of 0.44 nm and an angle of 80.4°.

Figure 1(c) presents the STM characterization of the P/Cu(513) system. The large-scale STM image on the left reveals high-quality, ultra-flat terraces of skewed-P with lateral dimensions of hundreds of nm. The atomically resolved STM image on the right shows the atomic structure of the P layer. Here, one can also notice the presence of sporadic adatoms and vacancies in the P lattice. To better solve the P structure, a close-up image is also shown as an inset, where the phosphorus unit cell is highlighted in black, with lattice parameters of $0.42 \times 0.42$ $nm^2$ and an angle of approx. 80°. This is in good agreement with a $1 \times 1$ P structure on Cu(513) facets, as illustrated by the Cu(513) model shown in Figure 1(b). Moreover, only one protrusion per unit cell is resolved in the STM image. This observation points to the presence of a single P atom per unit cell, as also supported by the simulated STM image shown in Figure S1 and the XPS measurement shown in Figure 1(d). Indeed, the P 2p core level deconvolution result in one main component located at binding energy of 128.35 eV, besides a much smaller contribution (peaks area ratio below 0.1) that is likely related to adatoms, vacancies, step edges, or other defects within the reconstructed P layer. An alternative interpretation relies on a shake-up feature associated with exciton formation. The latter would be consistent with the band structure analysis presented below, but further spectroscopic evidence would be needed to confirm it.

Next, the focus is turned to the position of the $1 \times 1$ P structure with respect to the substrate. The adsorption site of P atoms in the $1 \times 1$ structure was investigated by DFT calculations of a slab, in which we optimized the coordinates of a single P atom inside the substrate unit cell together with three underlying Cu atomic layers. As shown in Figure 1(b), we found the equilibrium P position slightly displaced from the bridge site between Cu surface atoms, such that P atoms protrude at the lowest possible height above the Cu(513) surface. Further details on the P–Cu distances are provided in Table ST1. To corroborate this model, we performed photoelectron diffraction (PED) measurements of the P 2p core level by varying the photon energy from 160 to 450 eV and recording the angular distribution at normal emission along the analyzer dispersion axis, in the $[\bar{3},0,1]$ direction. For each photon energy, the P 2p signal within a ±15° angular window was obtained, and the resulting intensity map is shown in the left-hand panel of Figure 1(e). The diffraction pattern arises from the interference of P photoelectron waves backscattered by both the Cu substrate and the P atoms of the overlayer, and therefore provides direct information on the atomic registry [48]. To validate the structural model, a photoelectron-diffraction map was simulated using first-principles calculations for different adsorption geometries of the P atoms. The computed map, shown in the right panel of Figure 1(e), reproduces the experimental features for both, photon energy and emission angle ranges. This excellent agreement provides strong and direct support for the proposed atomic structure of the new skewed-P phase. Moreover, additional PED simulations considering the possible incorporation of P atoms into the topmost Cu layer are presented in Figure S2. The simulations were performed for different percentages of P atoms substituting Cu atoms in the first atomic layer. As the fraction of incorporated P increases, the simulated PED patterns progressively deviate from the experimental results, with the characteristic features becoming less sharp and less well-defined. These results allow us to rule out a structure consisting exclusively of P atoms incorporated into the Cu substrate, forming a substitutional Cu–P alloy. Although the presence of a small fraction of incorporated P atoms cannot be entirely excluded, the PED simulations strongly support a structural model in which P atoms predominantly form an overlayer on the Cu surface. This interpretation is further supported by the XPS measurements, which reveal essentially a single P 2p component. In contrast, a significant fraction of P atoms incorporated into the Cu substrate would be expected to produce an additional chemically shifted component. Regarding a possible Cu–P overlayer alloy, such as the one proposed for the P/Cu(111) system [43], further investigation is required in order to safely exclude it, as such an alloy could be invisible to STM. Nevertheless, given the very short interlayer spacing of the (513) structure (see the side view in Fig. 1(b)), such Cu atoms would sit too close to the underlying Cu layer, leading to overlayer buckling, which is not observed experimentally.

Finally, we note that the nearest-neighbor P–P distance in this structure is 4.4 Å, which is too large to support conventional P–P bonding. Moreover, our calculations reveal a finite charge transfer from the Cu substrate to the P overlayer, resulting in a net accumulation of negative charge on the P atoms (see Table ST1), consistent with previous studies of P monolayers supported on Cu substrates [30,49] . Therefore, skewed-P is better described as a substrate-stabilized phosphorus atomic adlayer, rather than an independent 2D phosphorus allotrope with a continuous P–P bonding network.

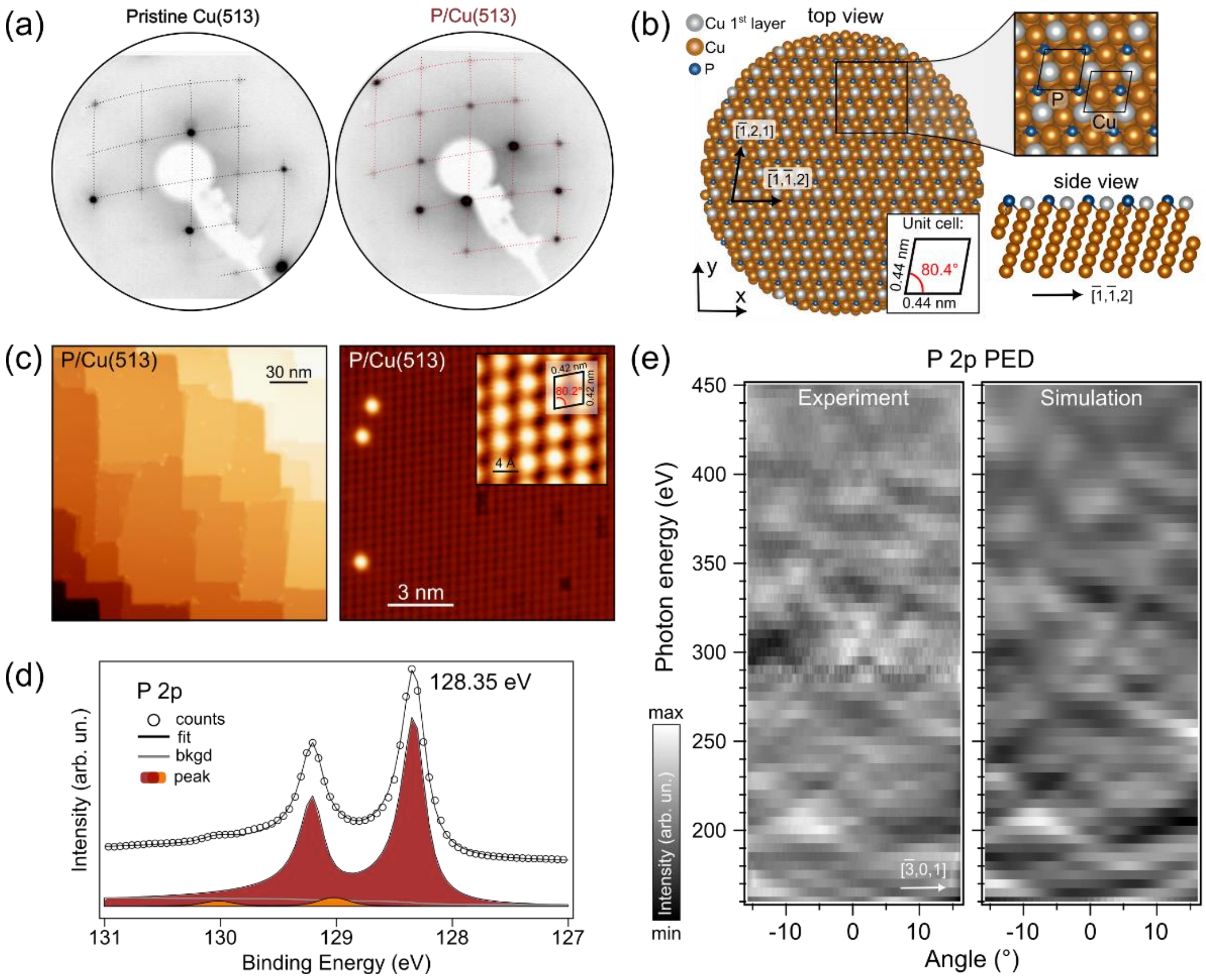

Figure 1. Structural characterization of P/Cu(513). a) LEED patterns taken at electron energy of 110 eV of the clean (left) and P-covered (right) Cu(513) surface, where the corresponding surface meshes are highlighted by black and red dotted lines, respectively. b) Structural model of the $1\times1$ P structure on Cu(513). On the right, zoomed images of the top and side views are shown, where the unit cells of Cu(513) and P are highlighted. c) Large-scale LT-STM image showing P/Cu(513) terraces. (Imaging parameters: I = 50 pA, V = 0.5 V, size = $200\times200$ nm$^2$) and an atomically resolved STM image of the P/Cu(513) system. (Imaging parameters: I = 300 pA, V = −0.5 V, size = $14\times14$ nm$^2$). In the inset, a close-up STM image is shown, where the P unit cell and its parameters are also indicated. (Imaging parameters: I = 300 pA, V = −0.5 V, size = $4\times4$ nm$^2$). d) P 2p XPS spectrum taken at photon energy of 250 eV. The best fit envelope and the deconvoluted peaks are included. e) Experimental (left) and simulated (right) photoelectron diffraction (PED) maps as a function of photon energy and emission angle along the $[\bar{3},0,1]$ direction at normal emission geometry.

The band structure of the P/Cu(513) system is thoroughly probed with angle-resolved photoemission spectroscopy, as shown across the different panels of Figure 2. In particular, Figure 2(a) shows a constant energy surface measured at a photon energy of 130 eV and cut at 1.2 eV below the Fermi level, which contains contributions from the bulk Cu sp-bands mainly. These bands are located in the upper and lower regions. Additionally, phosphorus-derived emissions are highlighted by arrows. The P bands can be divided into two main features, referred in the graph as 1 and 2. Feature 1 is centered around the $\bar{\Gamma}$ points, exhibiting a vertically oriented elliptical shape while feature 2 consists of pockets on the left-hand side of $\bar{\Gamma}$, symmetrically repeated on the right-hand side. The latter are characterized by a distorted rectangular shape that intersects the P unit cell edges (white lines). Notably, at the top of the image, the central feature 1 is repeated at a higher order $\bar{\Gamma}$ point, although only its upper portion is visible, whereas in the upper region the Cu band intensity obscures its observation.

To further characterize the phosphorus bands, Figures 2(b) and 2(c) show photoemission intensity mappings obtained from cuts through $\bar{\Gamma}$ along the vertical and horizontal directions, indicated by the green and magenta dashed lines in Figure 2(a), respectively. Considering the vertical cut of Figure 2(b), two Cu sp bands are observed at approximately $-1$ and 0.5 Å$^{-1}$, along with a P-related band (feature 1) exhibiting hole-like dispersion, with its maximum located at 0.8 eV below the Fermi level. A second, weaker hole-dispersing feature named 3 is also visible, centered at the $\bar{\Gamma}$ point with a maximum energy of approximately $-1.6$ eV. The horizontal cut shown in Figure 2(c) allows further investigation of the phosphorus-related feature 2. This cut is presented for two different light polarizations, whose geometries are sketched in Figure 1(e). The left panel shows linear horizontal (LH) polarization, which is also used for the constant energy surface in Figure 2(a) and the vertical cut in Figure 2(b). In this configuration, sketched on the left-hand side of Figure 2(e), the electric field of the incident light is 55° with respect to the surface plane and lies in the scattering plane. The right panel shows linear vertical (LV) polarization, where the electric field lies in the plane of the sample surface instead, as sketched on the right-hand side of Figure 2(e). In the LH configuration, all three main features are visible, with the central one showing enhanced intensity, while under LV polarization, the central feature drastically reduces its intensity, while the outer P bands reveal increased intensity. Additional energy dispersion plots obtained using left circular (CL) and right circular (CR) polarizations are shown and discussed in Figure S3.

Overall, none of the P-related bands in Figure 2 (a-c) crosses the Fermi level; the highest-energy states, associated with the outer features, lie approximately 0.8 eV below $E_F$. Since there are no further P-derived bands seen up to the Fermi level, this suggests the semiconducting-like character of the Phosphorus electronic states of the skewed-P overlayer. This interpretation is further supported by Figure S4, where the Scanning Tunneling Spectroscopy (STS) measured on P/Cu(513) is shown together with the calculated DOS, pointing to an estimated value of the band gap of 1.3 and 1.2 eV, respectively. However, these observation does not exclude possible admixture of the P-related states to the continuum of Cu states near the Fermi level. Thus, our finding does not imply a semiconducting behavior of the P/Cu(513) interface, since the underlying Cu substrate remains metallic, as evidenced, e.g., by the Fermi level crossing of the Cu s-p features. Rather, our work highlights how the substrate modifies the electronic structure of the P overlayer, with the observed P-induced bands remaining below the Fermi level for skewed-P on Cu(513) while crossing the Fermi level in the P/Cu(111) system.

We now discuss the bulk- versus surface-origin of the bands described above. To this end, we have measured the band dispersion along $k_x$ (at $k_y$=0 and $E-E_F=-0.9$ eV) as a function of photon energy, in the range 90–500 eV. The result is shown in Figure 2(d), which probes the energy dispersion along the out-of-plane ($k_z$) direction and allows distinction between bulk- and surface-derived bands. Two main types of features are visible: circular features (marked by black arrows) correspond to Cu sp-bulk bands, as they disperse along $k_z$, whereas the vertical lines originate from surface derived states, as the latter do not reveal dispersion along the out-of-plane direction. These features indicated by blue arrows are associated with the constant features arising from the P bands. Looking at Figure 2(c), at $E-E_F=-0.9$ eV, the profile intersects the apex of the three hole-like P bands previously discussed. These band maxima give rise, in Figure 2(d), to the three central vertical features. A fourth vertical line, visible at around 1.3 $Å^{-1}$, likely corresponds to a repetition of a P feature in the second Brillouin zone, which was not clearly observable at photon energy of 130 eV. To further support the assignment of features to surface and bulk states, Figure S5 shows the corresponding $k_z$-dependent data at the Fermi level. The absence of vertical features at this energy confirms the semiconducting-like behavior of the P-induced surface bands. Moreover, in Figure S6 we also provide additional mappings at 95 and 380 eV photon energies to better visualize the 2D character of the P derived bands in P/Cu(513).

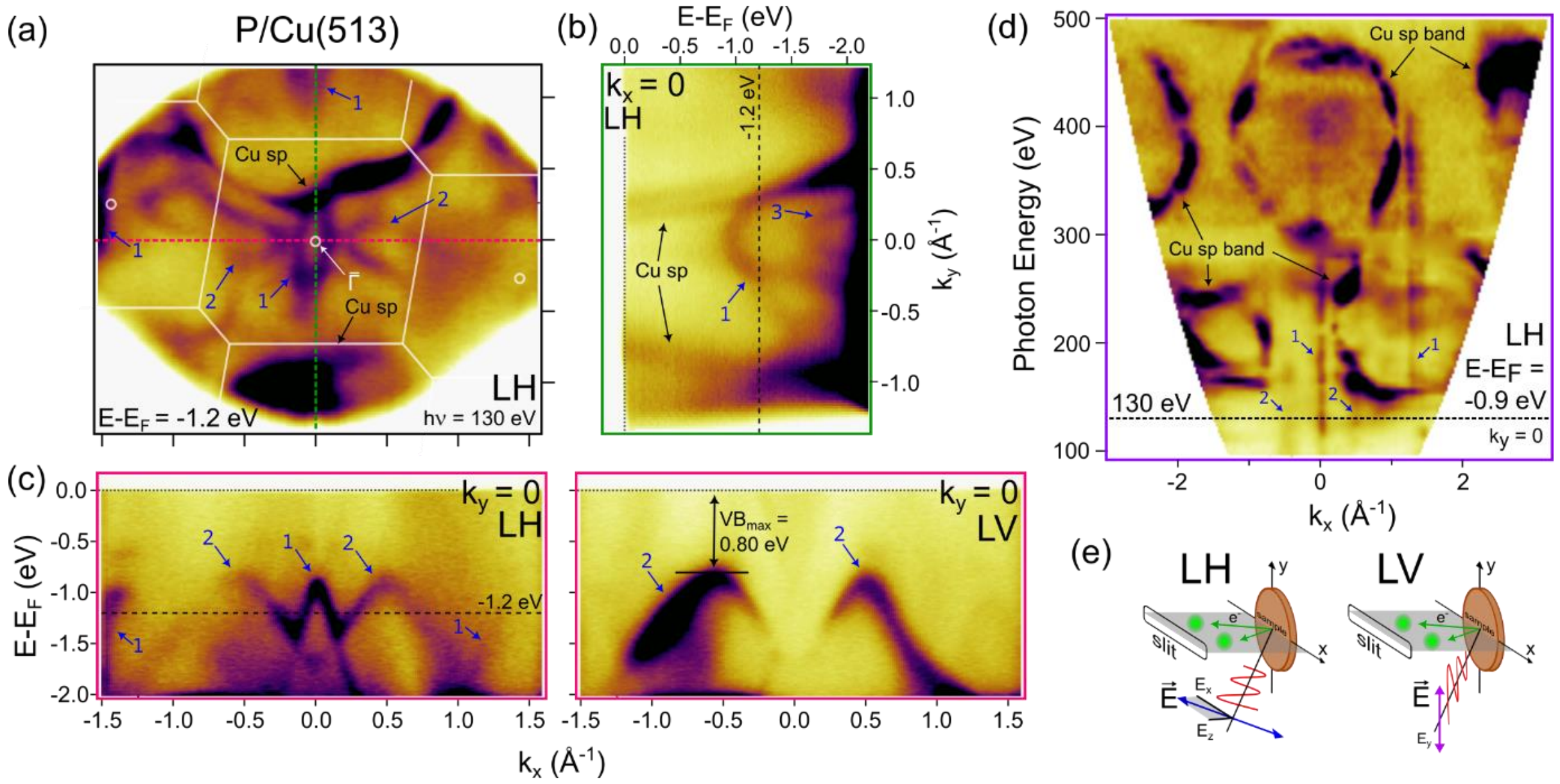

Figure 2. Angle resolved photoemission measurements on P/Cu(513). a) ARPES constant energy surface mapping of the P/Cu(513) system, taken at $E-E_F=-1.2$ eV with photon energy of 130 eV. The surface Brillouin zone of the 1 × 1 P structure is indicated in white. Black and blue arrows point to Cu- and P-derived bands,

respectively. The numbers label the P-related features. Green and magenta dashed lines correspond to the energy dispersion cuts presented in panels (b) and (c). b) Photoemission intensity mapping along the green dashed line in panel (a) ($k_x$=0). The energy used for the constant surface image in (a) is indicated here by the dashed black line, along with the Cu- and P-derived bands by the respective arrows. c) Photoemission intensity mappings along the magenta dashed line in panel (a) ($k_y$=0), measured using different light polarizations: linear horizontal (LH, left) and linear vertical (LV, right). d) Photoemission intensity mapping with LH polarization perpendicular to the surface ($k_z$) as a function of photon energy and wave vector $k_x$, measured at $k_y$=0 and shown here for $E-E_F = -0.9$ eV. Black and blue arrows indicate Cu- and P-derived features, respectively, while the numbers label the P bands again. e) Experimental geometry and polarizations used.

Further insight into the phosphorus-derived bands is obtained from DFT band structure calculations performed for the same optimized crystal structure that we used for PED modeling. To simulate the surface electronic band structure of the P/Cu(513) system, we performed LDA+U DFT calculation of a finite slab and extrapolated the DFT-derived electronic structure to obtain the spectral function for semi-infinite system (see the Method section for further details). The obtained spectral function for the surface region, which we defined as the P layer together with the four underlying Cu layers, is illustrated in Figure 3.

Figure 3(a) shows the simulated constant energy cuts of the surface spectral function to compare with the experimental photoemission intensity, shown in the inset. It is worth noting that all three P-related features observed experimentally are well reproduced by the simulation, indicating a very good agreement with the experiment. Moreover, the Cu sp band is nearly absent in the surface spectral function because it mostly belongs to bulk Cu. Figure 3(b) and 3(c) presents the simulated band dispersions along $k_x$=0 and $k_y$=0, respectively, allowing direct comparison with the analogous experimental data, also reported in the respective insets. The simulations again provide a very accurate description of the experimental features. For the $k_x$=0 dispersion shown in Figure 3(b), the band structure is similarly well reproduced by the simulation. The central P band shows minima at approximately −0.5 and +0.5 $Å^{-1}$ and a maximum at $k_y$=0. In the experimental data, a Cu sp band is additionally observed, likely resulting from an umklapp process, dispersing between −0.2 and −0.4 $Å^{-1}$. A second feature at $k_y$=0, visible up to binding energies of about −1.4 eV, is present in both the simulation and experiment, confirming its phosphorus origin rather than a copper d-band nature.

The DFT calculation for a hypothetical free-standing skewed-square P lattice yields a metallic band structure with a bandwidth of about 1.2 eV (see Figure S7). Placing the P layer on the Cu substrate shifts the P-derived bands. The slab DFT calculations reveal no additional well-defined P-related bands near $E_F$, in full agreement with the ARPES data.

At the same time, the PDOS of phosphorus remains nonzero near $E_F$ due to mixing of P orbitals with the metallic states of the Cu substrate. For this reason, we describe the absence of P-related bands at $E_F$ as a pseudogap, implying suppressed but nonzero density of P states near $E_F$. Thus, the adsorbed P atoms can be characterized as a quasi-semiconducting P-layer.

Focusing now on the $k_y$=0 dispersion of Figure 3(c), the main P-related bands 1 and 2 are clearly distinguishable, as indicated by arrows. The simulation also reveals an avoided crossing behavior when two P bands approach each other around −0.15 and +0.15 Å$^{-1}$, consistent with the experimental observations and indicative of hybridization between these bands. The LH/LV polarization dependence observed in the ARPES (see Figure 2(c)) directly matches the projected band-structure calculations result, which is shown in Figure 3(d). In particular, features 1 and 2 selectively appear in the $p_x$ and $p_y$ projections of the $k_x$=0 energy dispersion cut, respectively.

These bands are mostly composed of the in-plane oriented P 3$p_x$ and 3$p_y$ orbitals [50], as also indicated by the comparison between the in-plane ($p_x$ + $p_y$) and out-of-plane ($p_z$) P-orbital projections provided in Figure S8. On the other hand, the out-of-plane 3$p_z$ orbitals are strongly coupled to the Cu orbitals and therefore do not form well-defined bands; their contribution has the form of background in the spectral function. To further investigate the spatial localization of the surface-derived states, Figure S9 shows the Cu-layer-resolved band structure for the first five Cu layers below the P overlayer. The admixture of Cu orbitals to the surface-related bands is rapidly decreasing with depth and becomes small within the second to third Cu layers, confirming the localization of the states in the thin near-surface region.

Our experimental data can be qualitatively interpreted using the orbital selection rules for the dipole matrix element of photoemission that is governed by the following expression [51]:

$$\langle l'(\vec{k})|\vec{\varepsilon}\cdot\vec{r}|lm\rangle \propto \sum_{q,m'} \varepsilon_q C_{lm\,1q}^{l'm'} Y_{l'}^{m'}(\vec{k})$$

Here, $\vec{\varepsilon}$ is the polarization vector, $\vec{k}$ defines the photoelectron direction, $\varepsilon_{\pm1} = (\mp\varepsilon_x + i\varepsilon_y)/\sqrt{2}$, $\varepsilon_0 = \varepsilon_z$, $C$ stands for the Clebsch-Gordan coefficient and $Y_{l'}^{m'}$ is the spherical harmonic. From this, we obtain that for the LV polarization ($\varepsilon_{-1} = \varepsilon_1, \varepsilon_0 = 0$) the angular dependence of the photoemission matrix element from the $p_x$ orbital follows the $d_{xy}$ wave, leading to zero PE intensity in the $xz$ plane. However, for the $p_y$ orbital the matrix element is a combination of $s$, $d_{z^2}$ and $d_{x^2-y^2}$ waves, among which the first two functions give nonzero intensity in the $xz$ plane close to the z direction. For the LH polarization ($\varepsilon_{-1} = -\varepsilon_1, \varepsilon_0 \neq 0$), we find that for the $p_y$ orbital the matrix element is a combination of $d_{xy}$ and $d_{yz}$ waves that produce zero intensity in the $xz$ measurement plane. However, for the $p_x$ orbital the matrix element is a combination of $s$, $d_{z^2}$, $d_{xz}$ and $d_{x^2-y^2}$ waves that result in nonzero emission.

Accordingly, the LH geometry predominantly enhances bands with $p_x$ character, while the LV configuration selectively probes $p_y$-derived states, with $x$ and $y$ defined in the surface ($xy$) plane shown in the inset of Figure 3(d). Of course, our analysis neglects photoelectron diffraction that is a probable reason of nonzero intensity from $p_y$ states for the LH geometry. Nevertheless, on a qualitative level our interpretation reveals the strong correspondence between the polarization-resolved ARPES data and the calculated $p_x/p_y$ projections and, therefore, provides direct experimental confirmation of the orbital symmetry of the phosphorus-derived bands. As a benchmark, the band structure was also calculated using a GGA+U functional. The resulting dispersion, shown in Figure S10, closely reproduces the LDA+U result of Figure 3, with the main difference being an approximately rigid energy shift that places the valence-band maximum at 0.75 eV below $E_F$, compared with 0.80 eV for LDA+U. The overall band dispersion and relevant electronic features remain essentially unchanged, confirming the robustness of the calculation with respect to the exchange–correlation functional.

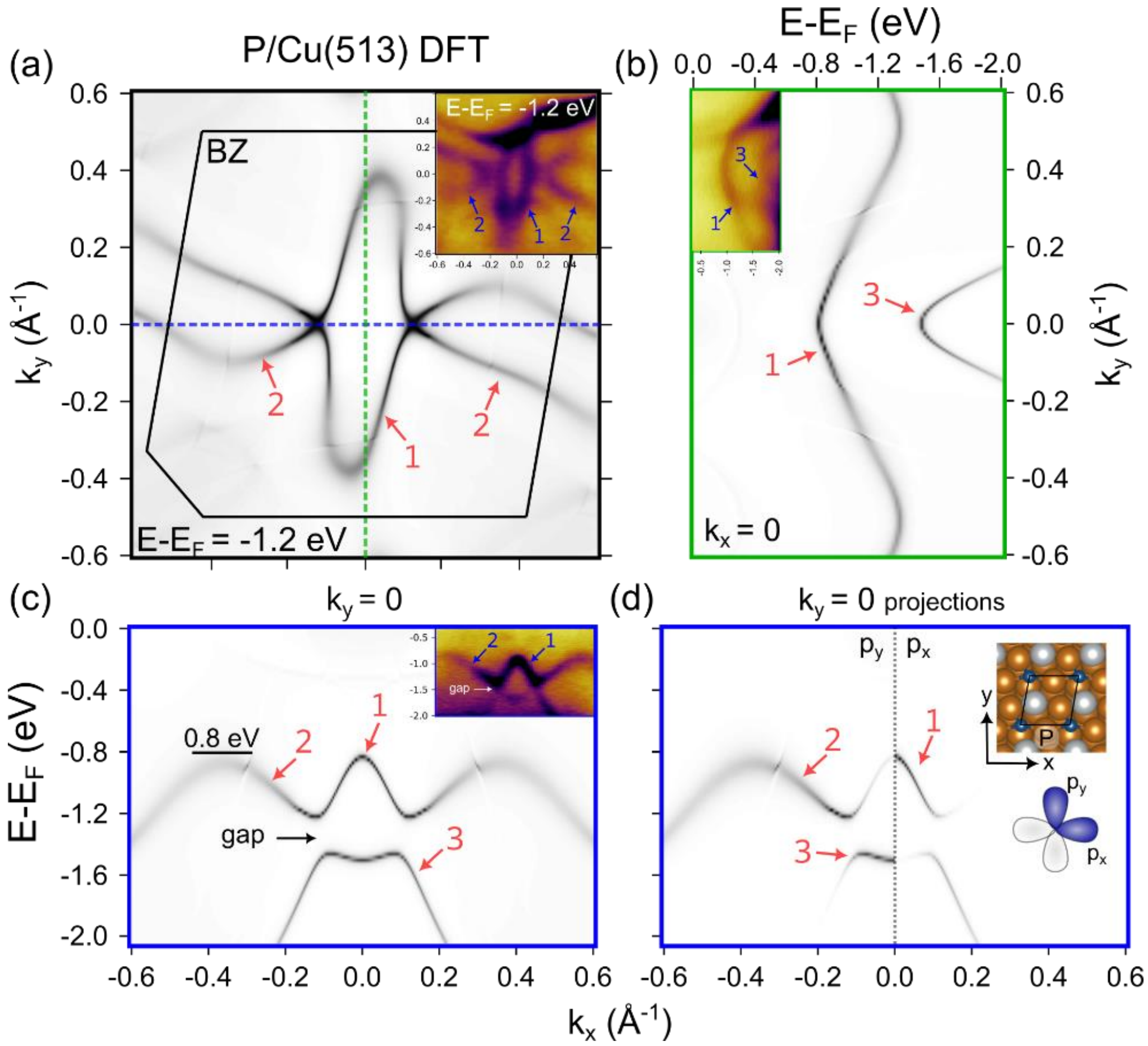

Figure 3. a) Simulated constant energy cut of the spectral function of the P/Cu(513) system, shown at $E-E_F = -1.2$ eV. Green and blue dashed lines indicate the energy dispersion cuts presented in panels (b-c). In the inset is shown the experimental constant energy cut at $E-E_F = -1.2$ eV for a direct comparison. b),c) Simulated energy dispersion plots along $k_x = 0$ and $k_y = 0$, as indicated by the green and blue dashed line in panel (a). In the insets, the analogous experimental measurement is shown. The maximum energy of the P bands is indicated. d) Projection of the $k_y$=0 simulated energy dispersion plot along $p_x$ (right half) and $p_y$ (left half). Inside the image are present the $p_x$ and $p_y$ orbitals oriented with respect to the P unit cell.

## 2. P-induced faceting on a curved Cu substrate.

Having fully characterized the structural and electronic properties of the Skewed-P phase, we now investigate its role as one of the two competing P phases that stabilize on the vicinal Cu(111) surfaces. To do so, we used a curved Cu crystal, referred to here as *c*-Cu(645), which exposes fully-kinked (111) vicinal surfaces, with the (645) direction located at its center, as sketched at the bottom of Figure 4(a). The curved crystal has a cylindrical geometry with a curvature radius of 16 mm. The sample curvature includes the (111) direction, for which the vicinal angle is defined as $\alpha = 0°$, and extends off-center to include higher-index planes within the crystal, reaching a vicinal angle of approximately 27° near the sample edge. The cylindrical shape allows measurement at a specific vicinal angle by simply moving the crystal laterally. The atomic steps exposed by this type of sample are of the fully-kinked type, characterized by a “zig-zag” atomic arrangement along the terrace edges, as previously described [52].

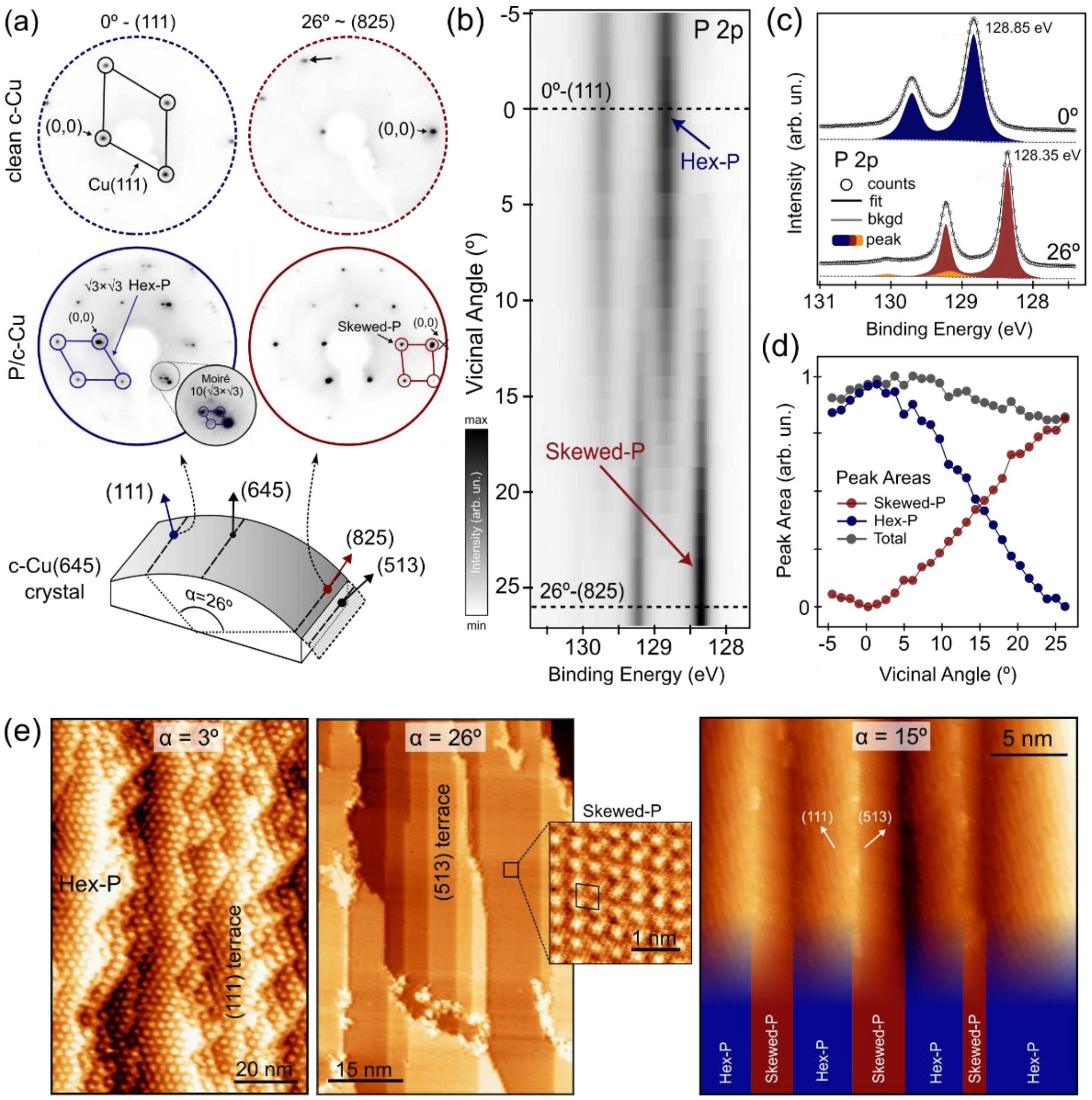

Figure 4. P-induced faceting on the *c*-Cu(645) substrate. (a) Top: LEED patterns at 92 eV electron energy of the clean *c*-Cu(645) sample acquired at vicinal angles $\alpha$ = 0° (left) and 26° (right), corresponding to the (111) and (825) directions, respectively. For 0°, the Cu(111) unit cell is outlined in black; for 26°, the arrow highlights the LEED spots with splitting. Middle: LEED patterns of the P/*c*-Cu(645) crystal at vicinal angles $\alpha$ = 0° (left) and 26° (right). At 0°, the hexagonal-P (Hex-P) unit cell is marked in blue; the inset shows a magnified view of the (0,0) spot, where the Moiré pattern is visible. At 26°, the unit cell of the skewed-P structure is marked in red together with its surface (0,0) center spot position. Bottom: Schematic description of the *c*-Cu(645) crystal used in the experiment. The relevant crystallographic directions are indicated, and black dotted arrows show the locations where the 0° and 26° LEED patterns were acquired. The grey extension illustrates the continuation of the sample curvature toward the (513) face, which is not included in the present *c*-Cu(645) sample by construction. (b) Photoemission intensity map of the P 2p core level as a function of vicinal angle.

Dashed lines highlight the relevant angles of 0° and 26°, while the arrows indicate the main features associated with hexagonal and skewed-P. (Photon energy 250 eV). (c) P 2p XPS spectra corresponding to 0° (top) and 26° (bottom) from the map shown in panel (b). The spectra are displayed together with the best-fit envelope, deconvoluted components, and background. (d) Peak areas of the 0° and 26° components shown in panel (b), calculated for each spectrum in the XPS map in panel (b), plotted as a function of vicinal angle together with their sum. (e) STM images taken at 3°, 26° and 15° vicinal angles. The inset shows a zoom of the (513) terrace. On the bottom of the right-hand side image, the sketch illustrates the self-defined nanoarray of P phases (Imaging parameters: I = 100 pA, V = −1.2 V).

In the top part of Figure 4(a), we show LEED patterns of *c*-Cu(645) acquired at two different vicinal angles, namely $\alpha = 0°$, corresponding to the (111) direction, and $\alpha = 26°$, located at the opposite edge of the crystal (left and right images, respectively), corresponding to the vicinity of the (825) direction. LEED patterns were recorded before and after phosphorus deposition (top and bottom images, respectively). The LEED pattern of the pristine Cu(111) surface exhibits the characteristic hexagonal lattice, although the (0,0) spot appears off-centered to the left due to the tilt of the (111) plane with respect to the electron beam that is perpendicular to the (645) surface at $\alpha = 9.2°$. At $\alpha = 26°$, the diffraction pattern deviates strongly from Cu(111) due to the fact that the Cu(111) terrace is not existing anymore, i.e., it has less than one atom. Below that limit, the LEED just reflect the atomic arrangement at the surface that is no longer hexagonal. There, at the (825) position, the (0,0) spot lies now on the opposite side of the image with respect to the (111) reflection, as the facet is now rotated by 26° relative to the (111).

Upon phosphorus deposition, the (111) side exhibits additional diffraction spots corresponding to the hexagonal-P structure, characterized by a $\sqrt{3} \times \sqrt{3}$ unit cell, together with a $10(\sqrt{3} \times \sqrt{3})$ Moiré superstructure (see zoomed inset), that arises from the lattice mismatch between the substrate and phosphorus, as previously reported [39]. Following the work of Kaddar *et al.*, our preparation shows the Moiré unit cell aligned with that of hexagonal-P, suggesting the formation of a single monolayer (1 ML) of P. In contrast, deposition beyond 1 ML leads to a Moiré unit cell aligned with the substrate lattice [39].

Focusing on the (825) side, the LEED pattern changes markedly, revealing a skewed-square structure with a side length of approximately 0.4 nm that matches with the one shown in Figure 1(a) and is ascribed to the skewed-P growing on the Cu(513) termination. In light of all above, P deposition induces significant faceting of the sample. Specifically, the Cu surface atoms reorganize in such a way as to accumulate atomic steps, leading to the formation of new terraces with different crystallographic orientations selected by phosphorus to better accommodate its stable structures. Apart from (111) terraces, the additional new terraces correspond to the (513) termination, where P grows as a 1×1 semiconducting-like phase, as explained above. Further confirmation arises from the position of the (0,0) center upon P deposition at 26°, which appears close to the border of the LEED screen at the opposite site as the (0,0) spot of the (111) surface. Figure S11 shows an integration of LEED images acquired over a wide range of electron energies, from which the evolution of the diffraction spots toward the new (0,0) center can be tracked. The (0,0) spots of the (111) and the new facet span a vicinal angle of 28°. This angle corresponds to the vicinal angle of (513) surface. It is worth to emphasize that the new (513) facet has a vicinal angle that is not originally present in the sample, as illustrated in the sketch at the bottom of Figure 4(a); therefore, it can only form as a result of this faceting process.

Qualitatively, faceting is favored if the elastic energy cost associated with redistributing Cu atoms to reshape its crystal surface is compensated by the increased stability of the P/Cu atomic interface formed on top. To emphasize this concept, phosphorus was also grown on a different curved Cu crystal, called *c*-Cu(111), characterized by straight steps and having the (111) direction at its center. As sketched in Figure S12, this second crystal has a cylindrical shape similar to that of *c*-Cu(645) but spans a range of crystallographic directions rotated by 90° with respect to *c*-Cu(645), including the (334)–(111)–(553) orientations, similar to the crystal described in previous work [53]. As shown by the LEED and STM images in Figure S13, acquired after P deposition, the P layer still induces strong (513)-faceting at high vicinal angles. However, in this case exposing the (513) facet requires terraces to reorient along the azimuthal direction, resulting in a "pyramidal" shaping of the local surface (see Supporting Information for details). Coming back to the P layer formation on the kinked step Cu(645) substrate in Figures 1-4, the formation of the skewed-P/Cu(513) phase was found to be reproducible over a broad range of preparation temperatures. As shown in Figure S14, the characteristic LEED pattern is already observed after P deposition with the substrate held at 290 °C and remains essentially unchanged after subsequent annealing at 410 °C for 10 min, indicating the structural robustness of the phase within this temperature range.

In Figure 4(b), a map of the P 2p XPS spectra of the P/*c*-Cu(645) system is shown as a function of the vicinal angle $\alpha$. The map reveals the coexistence of two main P 2p contributions across the sample. One feature reaches its maximum intensity at $\alpha = 0°$, corresponding to the hexagonal-P phase growing on the (111) terraces, while the second doublet exhibits maximum intensity at the edge of the crystal and is associated with the skewed-P. Figure 4(c) shows the P 2p spectra acquired at 0° and 26°, respectively. The P $2p_{3/2}$ peak measured at 0° is centered at binding energy of 128.85 eV, in good agreement with previous studies of hexagonal-P on Cu(111) [39]. It is worth nothing that the relatively large width of the fitted peak reflects the presence of a variety of P adsorption geometries in this two-dimensional material. Indeed, the lattice mismatch between hexagonal-P and Cu(111), which gives rise to the Moiré pattern observed in LEED, leads to P atoms occupying several distinct P–Cu registry sites, each contributing a slightly shifted P 2p component. Considering the P 2p peak measured at 26º, the optimal fit requires a very intense and sharp doublet at 128.35 eV ($2p_{3/2}$) and also a minor emission at 129.15 eV, again in excellent matching with the P 2p spectrum of the P/Cu(513) system and therefore confirming that there are two stable P phases competing on the substrate: the hexagonal-P and the Skewed-P, on the (111) and on the (513) terraces, respectively.

The evolution of the fitted peak intensities of all P 2p components, extracted from the spectral map in Figure 4(b), as a function of the vicinal angle is displayed in Figure 4(d). The intensity profile of the hexagonal-P phase (represented by the blue curve) reaches its maximum around $\alpha = 0°$, whereas the intensity profile of the skewed-P phase (depicted by the red curve) exhibits its minimum. Conversely, at the maximum vicinal angle of the crystal, i.e., at 26º, the intensity profile of the new P phase predominates, whereas the contribution of hexagonal-P becomes nearly negligible. Significantly, at a vicinal angle of $\alpha = 15°$, at Cu(423), the intensity profiles of both phases attain an inversion point. At this angle, both phases are equally existing, with each contributing 50% of the surface area. Regarding the intensities ratio, at Cu(513), the unit cell parameters (4.42 × 4.42 Å$^2$, 80.4º) yields a P atom density of 5.2 nm$^{-2}$ for the skewed-P structure, approx. 20% lower than the 6.6 nm$^{-2}$ density of P/Cu(111) [40]. This is consistent with the XPS intensity ratios when photoelectron diffraction effects are not taken into account [54]. The latter issue is justified by the small variation (15%) of P 2p XPS intensities taken at different photon energies, as indicated in Figure S15.

Locally, the P-induced faceting creates an alternation of nanometer-wide terraces stabilizing the two different P phases, as visible in the STM image shown at the bottom of Figure 4(e) and highlighted by the white lines. The image was acquired at an intermediate vicinal angle of 15º, nominally corresponding to the Cu(423) substrate orientation, where the relative widths of the two terminations are comparable. Closer to the (111) orientation, the hexagonal-P phase becomes predominant, whereas near the (513) orientation the terraces hosting the new P phase dominate, as shown in the STM images at the top left and top right part of Figure 1(e), taken at vicinal angle of approx. 3º and 26º, respectively. Due to the cylindrical geometry of the curved crystal, the terraces develop in a stripe-like fashion, with widths on the order of 5 nm. As illustrated in the sketch at the bottom of Figure 4(e), this results in a self-organized periodic alternation of metallic hexagonal-P and semiconducting-like skewed-P stripes, whose relative widths can be tuned through the vicinal angle. From a nanotechnology perspective, such an architecture may offer interesting possibilities for future device concepts. For example, the alternating metallic and semiconducting-like phosphorus regions could potentially serve as the basis for nanoscale transistor architectures, in which the metallic domains might act as naturally integrated contacts to the semiconducting regions. Since both phases originate from phosphorus, this may also favor the formation of low-resistance interfaces, although the actual transport properties of such junctions remain to be established. More generally, the fact that this architecture emerges spontaneously from P-induced faceting during a single red-phosphorus deposition step suggests a potentially scalable route to nanoscale structural and electronic patterning, making faceting-induced nanostructure engineering an attractive strategy for the future integration of 2D materials into device-oriented architectures.

## Conclusions:

In this work, we demonstrate that crystal-facet engineering on a curved Cu substrate enables the simultaneous stabilization of two structurally and electronically distinct phosphorus phases under identical growth conditions. A metallic phosphorous structure forms on Cu(111) terraces, whereas a previously unreported skewed-square overlayer, denoted skewed-P, is stabilized on Cu(513) facets and exhibits semiconducting-like behavior. Periodic faceting of the substrate therefore produces a self-organized nanoscale pattern of alternating phosphorus regions with distinct band dispersions, demonstrating that facet orientation can be used to tailor both the atomic arrangement and the electronic properties of the overlayer within a single growth process. More broadly, these results establish crystal-facet engineering as a route not only to discover and stabilize new substrate-supported phases on non-conventional surfaces, but may also generate electronically patterned surface nanoarchitectures through a simple single-step phosphorus deposition process.

## Methods:

*Sample Preparation.*

The *c*-Cu(645), c-Cu(111) and Cu(513) samples were prepared in Ultra-High Vacuum (UHV) conditions (pressure better than $10^{-7}$ Pa) by repeated cycles of sputtering ($Ar^+$, 10 mA, 1 kV) and annealing (500°C) until the Low Energy Electron Diffraction (LEED) was giving a sharp diffraction pattern. In the case of the curved samples, the LEED was checked all along the sample curvature, scanning the vicinal angle, by step of 0.5 mm. Red phosphorus (99.99% purity, Merck), which is demonstrated to yield the same P structures as black phosphorus [31,55] was sublimated from a Knudsen cell at a temperature of 320°C, as measured by a thermocouple attached to the cell, facing the clean samples kept at a temperature of 350°C. The formation of the correct phosphorus superstructure was checked by LEED on both samples.

*Experimental Techniques.*

All experiments were conducted in UHV conditions, with pressure better than $10^{-7}$ Pa. The X-Ray Photoemission Spectroscopy (XPS) (except for the spectrum shown in Figure 1(d)), Angle-Resolved Photoemission Spectroscopy (ARPES) and Photoelectron Diffraction (PED) were performed at LOREA beamline at ALBA synchrotron (Barcelona, Spain), where a MB Scientific A-1 momentum-resolving spectrometer with 0.2° angular resolution was employed. The XPS spectrum of Figure 1(d) were taken using $h\nu = 250$ eV. The XPS map of Figure 4(b) was performed by measuring the P 2p core level at photon energy $h\nu = 250$ eV every 0.5 mm along the sample curvature. The fit shown in Figure 4(c) and used for the peak areas plot of Figure 4(d) employed a Voigt line shape and subtraction of a Shirley background.

The XPS spectrum shown in Figure 1(d) was taken at GELEM beamline at BESSY II synchrotron (Berlin, Germany), using photon energy of $h\nu = 250$ eV. The fit employed a Voigt line shape and subtraction of a Shirley background.

The PED data were acquired by measuring the P 2p core level in angular-resolved mode at normal emission, ranging the photon energy from 160 to 450 eV. The intensity map of Figure 3(b) (experimental) was obtained by plotting the P 2p spectral intensity divided by the photon flux and by the power function of the photon energy $(h\nu)^{-n}$ with $n$ adjusted to eliminate the decrease of the P 2p photoemission cross section with increasing photon energy.

The STM data shown in Figure 1(b) were acquired on a low-temperature scanning tunnelling microscope (commercial ScientaOmicron, University of Oviedo, Mieres, Spain) with a chemically etched tungsten tip. The measurement temperature was 78 K. The STM data shown in Figure 4(e) were acquired at room temperature using a variable-temperature scanning tunnelling microscope (commercial ScientaOmicron, Centro de Física de Materiales, San Sebastián, Spain) with a chemically etched tungsten tip.

*Computational Details.*

Scalar relativistic DFT calculations were performed within the local density approximation (LDA) for the exchange–correlation potential using the Perdew–Wang parameterization (PW92), as implemented in the FPLO-18.00-52 code, an improved version of the original FPLO code [56,57]. To obtain the correct energy position of the Cu 3d states, we applied a Hubbard correction U in the fully localized limit. The optimal value of U, chosen from the comparison of the ARPES data with the calculated band structure, is 4 eV. Bulk crystals were modeled in the P1 space group. The Brillouin zone was sampled using a 12×12×1 k-mesh. The sample consisted of an asymmetric slab consisting of 25 atomic layers of Cu, with one P atom in the unit cell. The in-plane lattice parameter was fixed to 4.427 Å. Position of the P atomic layer and the next three layers of Cu were optimized until the forces reached $10^{-3}$ eV/Å. The band structure was modeled using a semi-infinite slab. For this purpose, we created a Wannier model of the bands using the same FPLO code with projective Wannier functions. To construct a projector, we used 3spd orbitals for P atoms, while for Cu we used 4sp and 3d orbitals. After constructing the model Hamiltonian, we calculated the surface and bulk spectral functions for a semi-infinite slab following the method described in Ref. [58].

PED was computed by means of the EDAC code [59]. The model cluster was generated based on the unit cell obtained from the DFT calculations. The inner potential was set to 7 eV. The Debye temperature was 343 K for Cu and 500 K for P. The cluster radius was set to $R_{max}$=22 Å, and we limited the orbital moment to $l_{max}$=6. The inelastic mean free path was calculated from the universal curve as $143/E^2 + 0.054 \cdot E^{0.5}$ [nm] with $E$ denoting the kinetic energy of photoelectrons in eV. The resulting PED pattern was smoothed to simulate the angular resolution of 1°. PED patters are presented after removal of the background by subtracting the average intensity first for each photon energy and then for each angle.

## Acknowledgments:

Part of the experiments were performed at LOREA beamline at ALBA Synchrotron with the collaboration of ALBA staff. LOREA was cofounded by the European Regional Development Fund (ERDF) within the Framework of the Smart Growth Operative Program 2014-2020.

Authors acknowledge the help of Jordi Prat for the invaluable technical support during beamtimes at LOREA and Marco Gobbi for the useful discussions.

We also thank the Helmholtz-Zentrum Berlin für Materialien und Energie for the allocation of synchrotron radiation beamtime at GELEM Dipole.

We acknowledge the financial support received from the Spanish MCIN/AEI/10.13039/501100011033 through grant PID2023-149158OB-C44 and PID2022-140845OB-C64, as well as the Basque Government through the grant IT-1591-22 and IT-2133-26. D.Yu.U. acknowledges Saint-Petersburg State University under research Project No. 125022702939-2.